\documentstyle[twocolumn,aps,prl,epsf,floats]{revtex}

 \def\beq{\begin{equation}}
 \def\eeq{\end{equation}}
 \def\beqa{\begin{eqnarray}}
 \def\eeqa{\end{eqnarray}}

\begin{document}
\draft

\preprint{NSF-ITP-97-070}

\title{Spatially resolved spectroscopy on a superconducting proximity nanostructure}

\author{M. Vinet, C. Chapelier and F. Lefloch }
\address{ D\'epartement de Recherche Fondamentale sur la Mati\`ere
Condens\'ee,\\SPSMS, CEA-Grenoble - 17 rue des Martyrs, 38054 Grenoble Cedex 9, France. }

\address{\bigskip\rm (April, 2000)}

\address{\parbox{14.5cm}{\rm\small
\bigskip
We investigated the local density of states (LDOS) of a normal metal (N) in good electrical contact with a superconductor (S) as a function of the distance $x$ to the NS interface. The sample consists of a pattern of alternate stripes of Au and Nb made by UV-lithography. We used a low temperature scanning tunneling microscope to record simultaneously $dI/dV(V,x)$ curves and the topographic profile $z(x)$. Nearby the NS interface,  all the  spectra show a dip near the Fermi energy but depending on the geometry, different behaviours can be distinguished. First, when the characteristic size of the normal metal $L$ is much larger than the coherence length $\xi_N \simeq \sqrt{\hbar D_N/ 2\Delta}$, the spectral extension of the dip decreases from $\Delta$ at the NS interface to zero at distances $x\gg\xi_N$. Second, when $L$ is comparable to $\xi_N$ the apparent gap in the LDOS is space-independent and related to the Thouless energy.
}}

\address{\parbox{14.5cm}{\bigskip \rm \small
\pacs{PACS numbers:
74.50.+r 
73.40.Gk
73.50.Bk
}}}
\maketitle

A normal metal in good metallic contact with a superconductor can acquire some superconducting properties and reciprocally the superconductivity can be affected by the normal metal vicinity. This phenomenon is known as the proximity effect. It has been first extensively studied in the late 1960's using the Ginzburg-Landau theory which describes macroscopic superconducting systems near their transition temperature $T_c$ \cite{deGennes,Arnold}. Recent technical and material developments renewed the interest in the physics of this effect, especially at the mesoscopic scale. Simultaneously a more comprehensive
understanding of the proximity effect in the diffusive regime based
on the theory of non-equilibrium superconductivity has emerged \cite{bruder,Golubov}. For instance, predictions on the spatial dependence of the local density of states (LDOS) of a proximity structure are made. At very low temperature, when the
characteristic size $L$ of the normal metal becomes smaller than the
thermal length $L_T=\sqrt{\hbar D_N/2\pi k_B T}$ ($D_N$ is the
diffusion constant in the normal metal, $k_B$ the Boltzman constant, $\Delta$ the superconducting gap), the variation of the LDOS of a NS structure is predicted to depend on the ratio $L/\xi_N$ with $\xi_N=\sqrt{\hbar D_N/2\Delta}$ the
coherence length \cite{belzig,Golubov}. When $L\gg\xi_N$, the superconducting correlations 
extend to distances larger than $\xi_N$ leading to a depression of the
electronic density of states around $E_F$, the Fermi energy. Experimental tests of this theory were obtained by Gu\'eron \emph{et al.} \cite{Gueron}. They used nanofabricated tunnel junctions to probe the electronic density of states $n(E,x)$ at three distinct positions along a  normal wire in contact with a superconductor.  Scanning tunneling microscopy (STM) was also used to study this proximity effects in ballistic N metals (\emph{i.e.} $L\ll\xi_N,l_N$ where $l_N$ is the elastic mean free path in N) \cite{Tessmer,Truscott}. In these experiments the induced gap crucially depends on the thickness of the normal system.
The shape of the spectra is explained within the de Gennes-Saint James bound states model \cite{degennes}. In a diffusive sample where $l_N\ll L \leq \xi_N$, a gap  whose value depends on the Thouless energy, $E_{Th}=\hbar D/L^2$, is predicted \cite{belzig,pilgram}. STM experiments on small N wires embedded in a S matrix showed a healing length of superconductivity on the S side much larger than the coherence length $\xi_S$. But according to the authors, they could not  easily interpret the spatial dependence of the LDOS in the N metal because of a complicated geometry.\newline
In this Letter we report measurements of the local electronic density of states (LDOS) by scanning tunneling spectroscopy on Nb/Au proximity junctions. We spatially resolve the LDOS in the normal metal as a function of the distance $x$ to the NS interface. We  can discriminate between the two situations: $L \leq \xi_N$ and $L\gg \xi_N$ within the same sample. This rules out any spurious effects due to modifications of the film properties and interface qualities.

Spectroscopy by STM allows a high energetic and spatial resolution in conjonction with sample morphology. The STM hangs inside a sealed tube by a one meter long spring to decouple it from external vibrations. This tube is immersed in a He$^4$ cryostat. Cooling is achieved by introducing high purity helium exchange gas into the tube. The temperature is reduced to 1.5\,K by pumping on the helium bath. Tunneling spectra are obtained with a lock-in detection technique with a 77\,$\mu$V peak-to-peak modulation voltage of the bias at 1\,kHz. We measure the differential conductance $dI/dV$ versus $V$ while holding the STM tip at a fixed height above each position {\bf r}. This provides a local probe of $n(eV,{\bf r})$. The energetic resolution is determined by the thermal broadening at 1.5\,K. Each curve is a single acquisition process with no additional averaging. The $dI/dV$ curves and the topography are measured during the same line scan allowing us to correlate them  precisely  with the distance  $x$ to the NS interface. Data are normalized to the conductance at high voltage ($V>\Delta$). The \emph{same} normalization  coefficient has been used for all the curves. The typical tunneling resistance is $10^7\,\Omega$.

\begin{figure}
\epsfxsize=1.\hsize
\centerline{ \epsffile{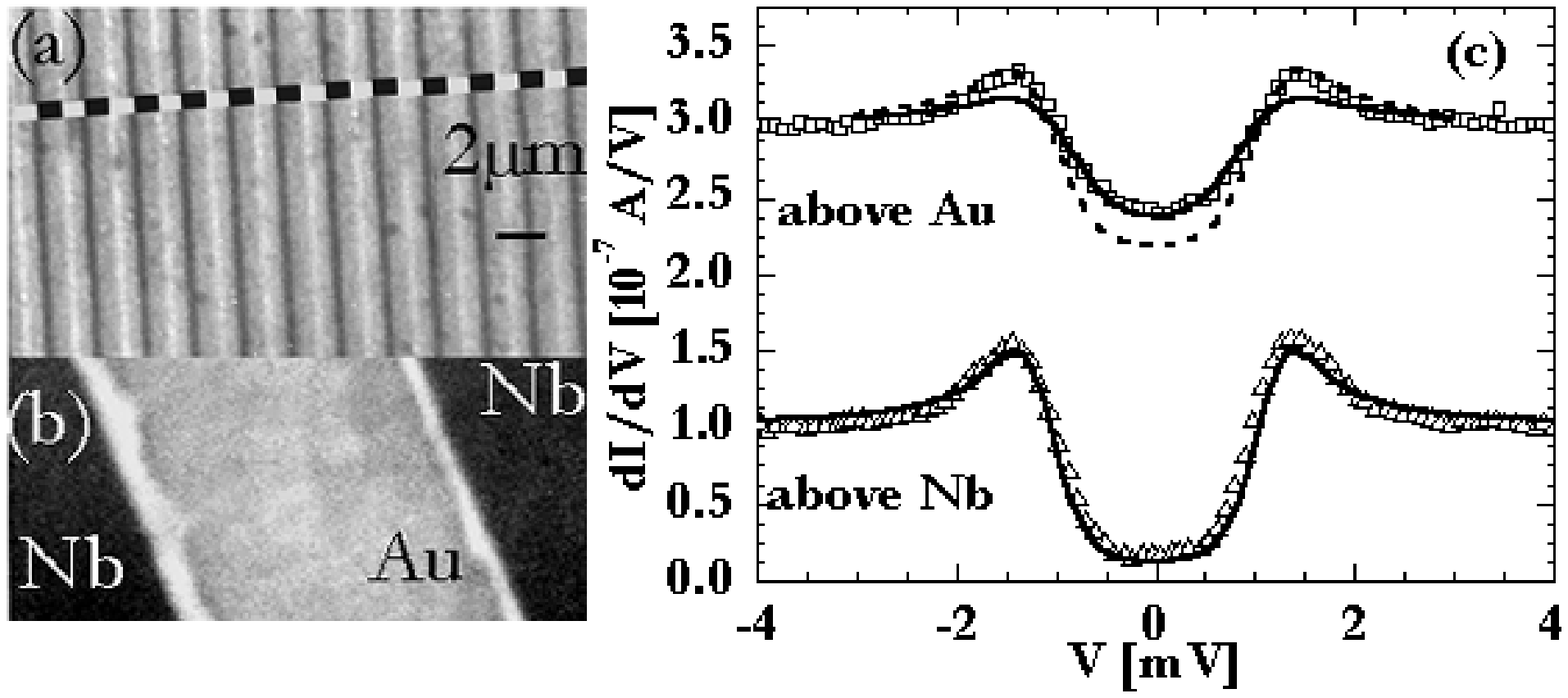}}
\refstepcounter{figure}
\label{photo}
FIG.\ref{photo}: 
{\rm\small
(a) Optical microscope picture of the sample. It consists of a 50 nm thick pattern of self aligned stripes 1\,$\mu$m wide and 250\,$\mu$m long of Au and Nb. The line represents section of the film. Inset (b) is a scanning electron microscope photography of the sample displaying the small ridges on each side of the Au stripe. Part (c) is a set of two spectra taken above  either Au or Nb. The experimental curves have been shifted for clarity. Above the Nb film (lower curve) the spectrum is fitted within a BCS model with $\Delta=1.15$\,meV and $\Gamma=0.07\Delta$ (solid line). In contrast spectra taken above Au (upper curve) can not be fitted within this model. The fit cannot describe simultaneously the wings and the zero-bias conductance. The dotted line is a fit with $\Delta=1.1$\,meV, $\Gamma=0.12\Delta$ whereas the solid one is for $\Delta=1.1$\,meV, $\Gamma=0.40\Delta$
}
\end{figure}

We have chosen Au as a normal metal because it is chemically inert. The Au film has been previously characterized by measuring the temperature dependence of the resistivity, $R(T)$, which gives a mean free path $l_N\simeq 22$\,nm and a diffusion coefficient $D_N\simeq 1.0\times10^{-2}$\,m$^2$s$^{-1}$ leading to $\xi_N\simeq 53$\,nm. Nb is used as a superconductor because of its high critical temperature. A 50\,nm thick Nb film  undergoes a BCS transition at $8.1$\,K and shows an energy gap $\Delta \simeq 1.15$\,meV at 1.5\,K. Its coherence length is $\xi_S=\sqrt{\hbar D_S/2\Delta}\simeq 27$\,nm and the mean free path is $l_S\simeq 6$\,nm given by previous $R(T)$ measurements. The ratio $l_S/\xi_S<1$ is characteristic of a dirty superconductor. Our sample consists of a 50\,nm thick pattern of juxtaposed stripes, 1\,$\mu$m wide and 250\,$\mu$m long, of Au and Nb in good electrical contact, see fig.\,\ref{photo}(a) and \ref{photo}(b).

 The Nb was first uniformly deposited by DC sputtering. We used UV-lithography and subsequent reactive ion etching (RIE) to produce the Nb stripes. We then sputtered 2\,nm Ti and 50\,nm Au and lifted-off the remaining resist. As seen in figure\,\ref{photo}(c), Au forms ridges on each side of the Nb stripes. The size of these ridges has been measured both by STM and scanning electron microscopy. Their lateral dimensions vary from 50\,nm to 200\,nm and their height from 10\,nm to 50\,nm. They are in electrical contact with Nb but poorly connected to the rest of Au. Indeed we observe BCS-like $dI/dV(V)$ curves in between the Au ridges and the Au stripes which indicates the presence of bare Nb (see lower curve of fig\,\ref{photo}(c)). The solid line represents the convolution of the density of states with a Fermi-Dirac distribution at the measured temperature $T=1.5$\,K. We introduce a phenomenological parameter $\Gamma$ to define a complex energy $E^* = E-i\Gamma$ \cite{btk}. This parameter reflects the broadening of the energy levels due to the finite lifetime of the quasi-particles. The calculation provides a superconducting gap $\Delta=1.15$ meV and a finite lifetime of the quasi-particles $\Gamma=0.07\Delta$. In contrast spectra taken above Au either over the ridges or the stripes can not be fitted using this model. As an illustration, the upper part of fig.\,\ref{photo}(c) shows several curves with different sets of parameters. The transition from these two types of spectra allows us to infer the NS interface.

In order to describe our results we need to take into account coherence effects of the quasi-particles inside the normal metal. In the framework of the quasi-classical Green's function formalism, induced correlations between electrons of opposite spin in the normal metal are described by a complex function $\theta (E,{\bf r})$ \cite{belzig,Esteve}. The LDOS of the quasi-particles $n$ is related to $\theta$ by $n(E,{\bf r})=n_0Re[\cos\theta (E,{\bf r})] $. In the dirty limit and in zero magnetic field, $\theta(E,{\bf r})$ obeys the 1D-diffusion Usadel equation \cite{usadel}:
\begin{eqnarray}
\frac{\hbar D}{2}\frac{\partial^2\theta}{\partial x^2} +[(-iE+\Gamma_{in})-2\Gamma_{sf}\cos\theta]\sin\theta & &\nonumber\\
+\Delta(x)\cos\theta &= &0, \label{green}
\end{eqnarray}
where $\Gamma_{sf}$ and  $\Gamma_{in}$ are the spin-flip and  the inelastic scattering rates respectively, $\Delta( x)$  the pair potential. In the normal metal ($x>0$) we assume no pair potential  ($\Delta_N=0$) whereas in the superconductor ($x<0$), $\Delta( x)$ obeys a self-consistent equation. For simplicity, we take $\Delta$  equals to its bulk value everywhere in the superconductor.
Conditions of continuity, $\theta_{x=0^-}=\theta_{x=0^+}$ and  spectral current conservation $\sigma_S(\frac{\partial\theta}{\partial x})_{x=0^-}=\sigma_N(\frac{\partial\theta}{\partial x})_{x=0^+}$ are imposed at the NS interface in the case of a perfect electrical contact.
\begin{figure}[t,\clip]
\epsfxsize=1\hsize
\centerline{ \epsffile{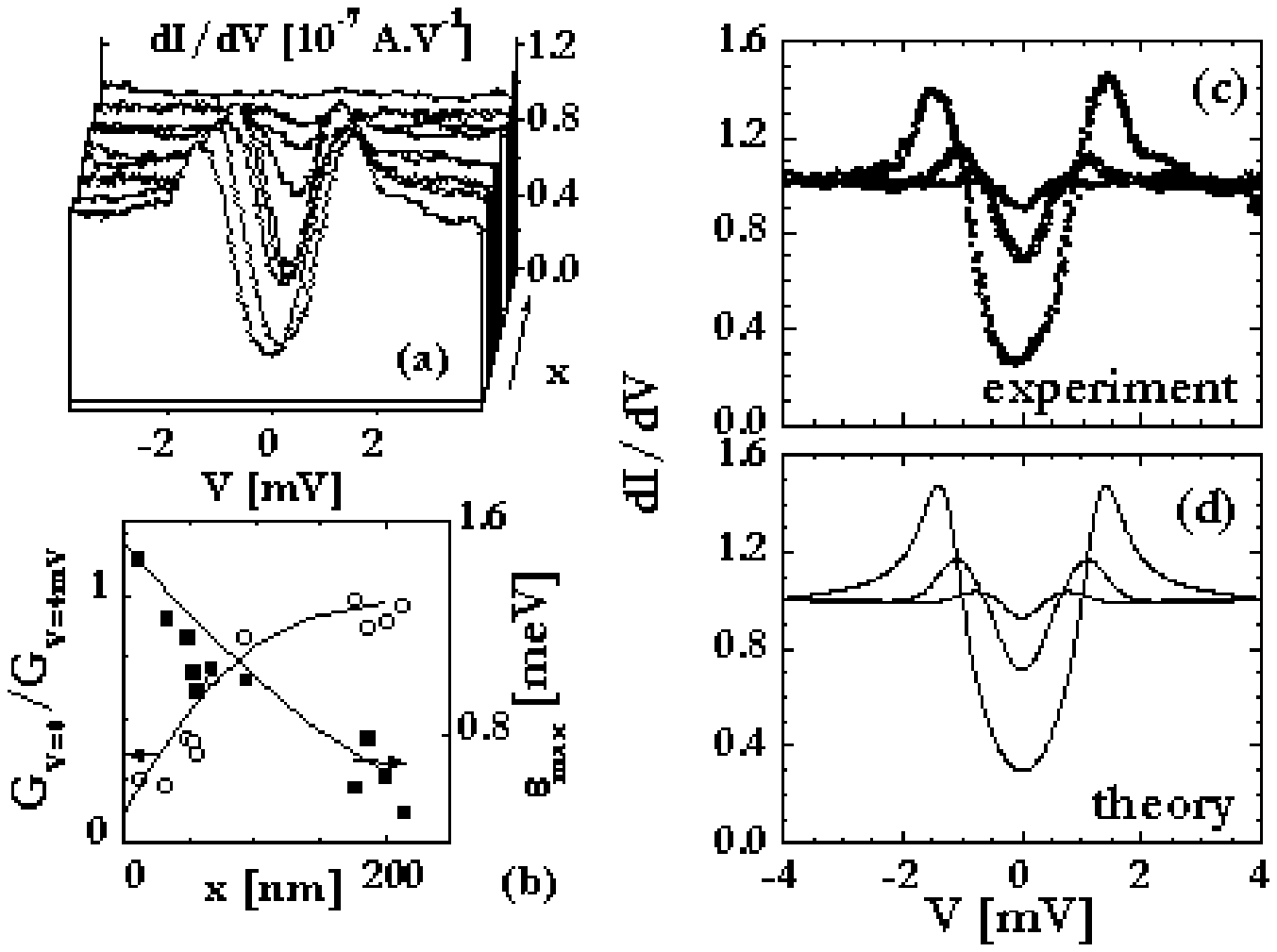} }
\refstepcounter{figure}
\label{infini}
FIG.\ref{infini}: 
{\rm\small(a) 3D plot of $dI/dV$ versus $V$ curves for different positions of the tip above the wide stripe of Au. The curves are evenly shifted along the $x$ axis (not to scale). The  pseudo-gap decreases from $\Delta$ at the NS interface to $0$ at distances larger than $200$ nm. (b) represents the normalized zero-bias conductance (open circles) and the energy of the peak (full squares), $\varepsilon_{max}$, as a function of the distance to the NS interface. The solid  lines correspond to the fit. On the right hand side, we compare measured spectra (c) with those calculated (d) for three distances: 11\,nm, 68\,nm and 185\,nm.}
\end{figure}

When $L\gg\xi_N$, the normal metal can be considered as semi-infinite and one recovers the N density of states far away from the superconductor. The boundary condition of eq.(\ref{green}) on the N side is then $\theta (E,\infty)=\theta_N=0$. The resulting LDOS $n(E,x)$ is predicted to display a space-dependent pseudo-gap structure \cite{belzig}. This situation corresponds to the 1\,$\mu$m wide stripe of Au. In the first set of experiments we scanned above its surface and measured the LDOS as a function of the distance $x$ to the NS interface. The spectra were taken along the same line perpendicular to the interface. The space dependent LDOS is plotted in fig.\,\ref{infini}(a) for different $x$.

We observe a peak in $G_V(x)=dI/dV(V,x)$ at an energy $\varepsilon_{max}$ which decreases as we move away from the superconductor.  Figure\,\ref{infini}(b) displays $\varepsilon_{max}$ and the normalized zero bias conductance $G_{V=0}/G_{V=4mV}$ as a function of $x$. A zero density of states at the Fermi level is never observed due to pair breaking mechanisms such as spin-flip and inelastic scattering which cannot be neglected at 1.5\,K. 
We compare our measurements to the convolution of the LDOS calculated from eq.\,(\ref{green}) with a Fermi-Dirac distribution. For the superconducting part, we used the value of $\Delta$ and $\Gamma_{in}$ given by the BCS fit above Nb. For the normal part, the rates $\Gamma_{in}$, $\Gamma_{sf}$ and the ratios $\xi_S/\xi_N$, $\sigma_S/\sigma_N$ are taken as adjustable parameters. We find  $\Gamma_{in}=0.01\Delta$, $\Gamma_{sf}=0.15\Delta$ and the ratios $\xi_S/\xi_N=0.2$, $\sigma_S/\sigma_N=4$. The deduced mismatch parameter $\gamma=\sigma_N\xi_S/\sigma_S\xi_N$ equals 0.05 instead of 2.5 evaluated from the conductance measurements obtained on films on the same thicknesses. This discrepancy reflects a much stronger induced superconductivity in the normal metal than expected \cite{golubov2}. This effect may be due to intermixing between Au and Nb at the interface or to the presence of Ti in between these two materials. Indeed the transport measurements do not provide any information on the influence of Ti on the conductivity of  Au. Moreover from $\Gamma_{in}$ and $\Gamma_{sf}$ we can deduce a decoherence time $\tau=h/max(\Gamma_{in},\Gamma_{sf})\simeq 20$\,ps. This value is in good agreement with recent experiments measuring the phase relaxation time on Au films deposited above  either Ti \cite{baselmans} or Al \cite{pothier}.

The fluctuations of $G_0$ and $\varepsilon_{max}$  are not due to experimental noise but are correlated to changes in the Au film morphology related to different grains.  A slight dispersion of the properties of these grains may explain such non-monotic evolutions \cite{Truscott}. 

\begin{figure}[t]
\epsfxsize=1 \hsize
\centerline{ \epsffile{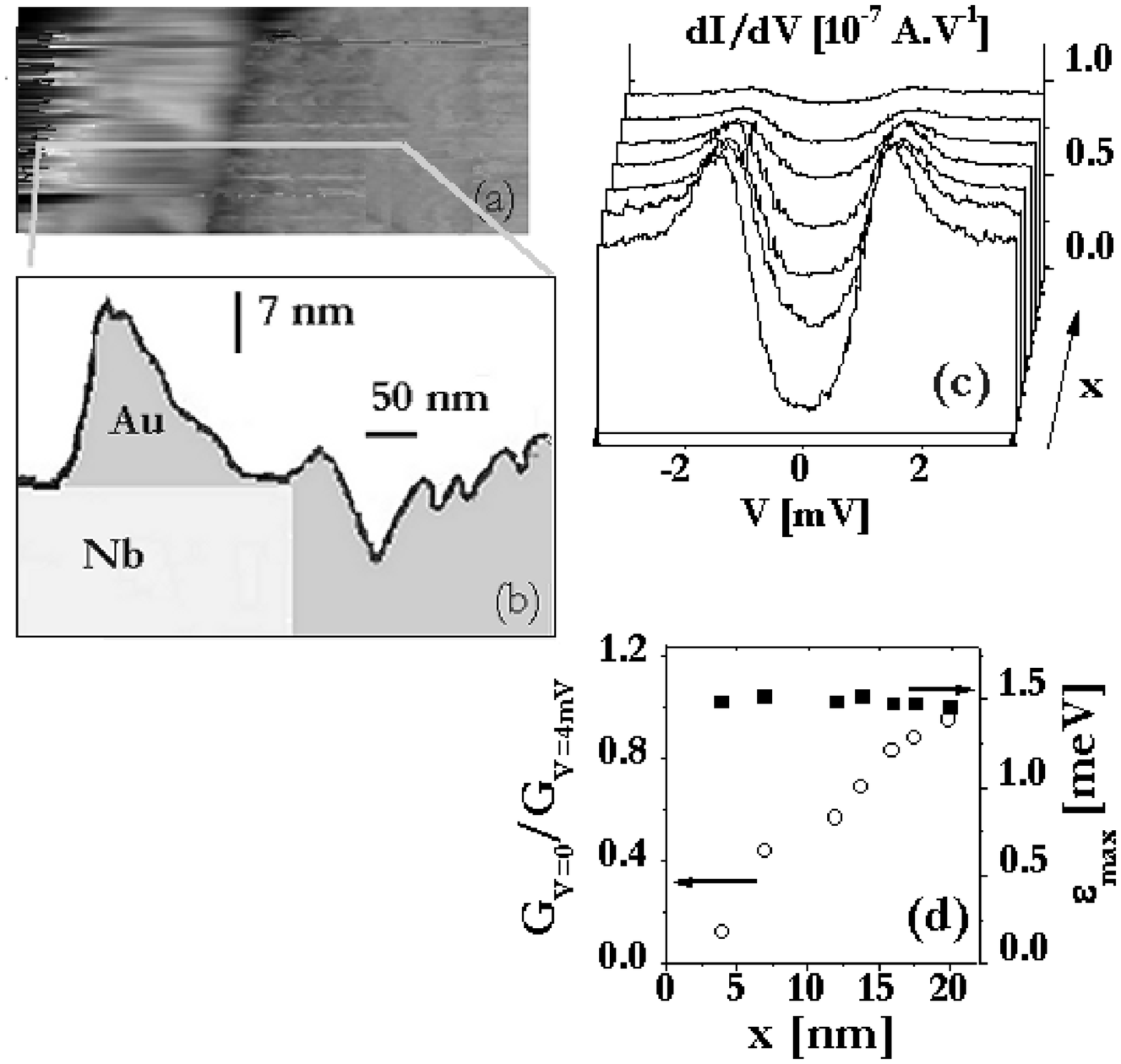}}
\refstepcounter{figure}
\label{petitegoutte}
FIG.\ref{petitegoutte}: 
{\rm\small
(a) shows a 400\,nm$\times$400\,nm STM topographic image of a ridge 20\,nm high. (b)is a measured line profile. (c) the 3D plots represents spectra for different positions above the Au ridge. (d) the zero bias conductance, $G_{V=0}/G_{V=4mV}$, and the energy of the peaks, $\varepsilon_{max}$ are plotted as a function of the distance $x$ to the NS interface. $\varepsilon_{max}=1.5$\,meV does not vary  while $G_{V=0}/G_{V=4mV}$ continuously increases with $x$.
}
\end{figure}

\begin{figure}[t]
\epsfxsize=1 \hsize
\centerline{ \epsffile{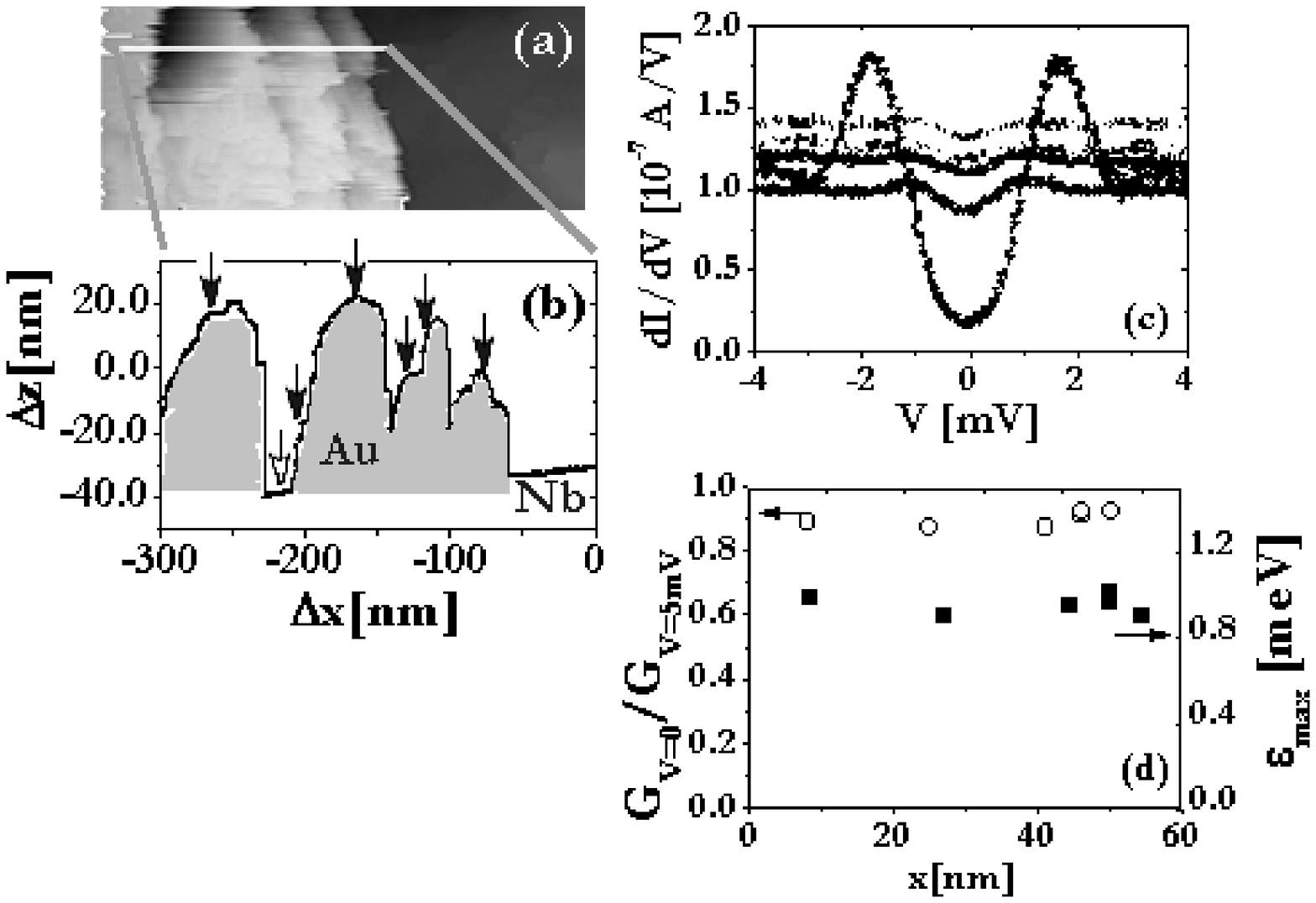}}
\refstepcounter{figure}
\label{grossegoutte}
FIG.\ref{grossegoutte}: 
{\rm\small
Picture (a) represents a STM topographic image of 550\,nm$\times$550\,nm, we have plotted a profile of the measured ridge. Spectra have been taken above this ridge along the white line in (a). The arrows represents places where spectra have been taken. For clarity  $dI/dV(v)$ curves taken from left to right have been shifted in (c). The ones correspponding to the black arrows are all similar except for the one corresponding to the white arrow taken in the bottom of the ridge. We have plotted $\varepsilon_{max}$ and $G_{V=0}/G_{V=4mV}$ for all the similar spectra as a function of $x$ in (d) to prove that they are space independent.
}
\end{figure}

In the second step of experiments, we have investigated the LDOS over two small ridges of Au. They are lying over Nb and the NS interface is now horizontal. Due to the initial slope of their edges, when moving the STM tip horizontally we can change the distance to  the NS interface (see figures\,\ref{petitegoutte} and \ref{grossegoutte}). Unlike previous experiments where the proximity effect was probed on NS bilayers with different thicknesses  \cite{Truscott}, we do not change the overall size of the normal system between different spectra but only the distance to the NS interface. Therefore we obtain the actual spatial dependence of the LDOS $n(E,x)$ of a normal system with a fixed dimension $L$.

In this configuration, we find that the position of the peaks in $dI/dV(V,x)$ does not depend on the distance to the NS interface in striking contrast to the preceding case. For a ridge of 20\,nm height, we observe that the maximum is located at 1.5\,meV (see fig.\,\ref{petitegoutte}) whereas in a ridge of height 50\,nm, the position of the peak is 0.95\,meV (see fig.\,\ref{grossegoutte}). Theoretically, in such a finite diffusive geometry \emph{i.e.} when $L$ is of the order of $\xi_N$, the N side boundary condition of eq.\,(\ref{green}) is $\frac{\partial\theta}{\partial x}(E,x=L)=0$ and a minigap $E_g$ is predicted to appear in the LDOS. $E_g$ is related to the Thouless energy $E_{Th}=\hbar D_N/L^{2}$ and $E_g=min(\Delta,\Delta(1+L/2.1\xi_N)^{-2})$ \cite{belzig,Golubov}. In the first ridge, $E_{Th}=18$\,meV$\gg\Delta$, the minigap is limited by the superconducting one. On the contrary, in the second ridge, $E_{Th}=2.8$\,meV that yields to a minigap $E_g$=0.53\,meV. Eventhough, we measure the position of the peaks and not the value of the minigap, our results are consistent with the expected behaviour of the gap as a function of the size $L$. Further numerical simulations are still needed to get a more quantitative comparison.

The evolution of the zero bias conductance with $x$ is different for the two ridges. In the bigger one, the LDOS suddenly changes as soon as the tip probes the Au surface and remains identical for any location above the ridge  (fig.\,\ref{grossegoutte})  whereas in the smaller one, $G_{V=0}/G_{V=4mV}$ continuously increases with $x$ from 0.1 to 0.9 (fig.\,\ref{petitegoutte}). This can be due to 3D geometrical effects since in this latter case the normal part is in a quasi-ballistic regime ($L\simeq l_N$).

In conclusion, we have probed the proximity effect in two different systems. First, we have investigated the LDOS of a semi-infinite normal metal and found space dependent energy spectra as a function of the distance to the NS interface. This behaviour is in good agreement with the pseudo-gap model predicted by the theory of non-equilibrium superconductivity. We have also investigated the local density of states in a confined geometry. In this case we have found a spectral structure which does not vary in space and which can be related to the Thouless energy. However the evolution of the zero bias conductance depends on the shape of the normal metal.

We wish to acknowledge beneficial discussions with M. Sanquer and the valuable help of J.C. Toussaint for numerical calculations.

\end{document}